\documentclass[runningheads,table]{llncs}
\newcommand{\doctitle}{Recursive Variable-Length State Compression}
\newcommand{\docsubject}{for Multi-Core Software Model Checking}
\newcommand{\docauthor}{Freark I. van der Berg}
\newcommand{\docauthorrunning}{F. I. van der Berg}
\newcommand{\docaff}{\emph{Formal Methods and Tools, University of Twente}, Enschede, The Netherlands}
\newcommand{\docemail}{f.i.vanderberg@utwente.nl}

\usepackage{hyperref}
\usepackage{amsmath,amsfonts}
\usepackage{algorithmic}
\usepackage{wrapfig}
\usepackage{xspace}
\usepackage{listings}
\usepackage[final]{pdfpages}
\usepackage{pdfcomment}
\usepackage{pstricks}
\usepackage{xspace}
\usepackage{svg}
\usepackage{cleveref}
\usepackage{fixfoot}
\usepackage[percent]{overpic}
\usepackage{marginnote}
\usepackage{suffix}
\usepackage{float}
\usepackage{multirow}
\usepackage[absolute]{textpos}
\usepackage{rotating}
\usepackage{flafter}
\usepackage{afterpage}
\usepackage{tikz}

\usepackage{subcaption}

\usepackage{xparse}
\usepackage{pdflscape}
\usepackage{graphbox}
\usepackage{enumitem}

\newcommand{\code}[1]{\texttt{#1}}%

\newcommand{\casweakcode}{$\mathtt{CAS_{weak}}$}%
\newcommand{\casstrongcode}{$\mathtt{CAS_{strong}}$}%

\hypersetup{
pdfauthor={\docauthor},
pdftitle={\doctitle},
pdfsubject={\docsubject},
colorlinks=true,
citecolor=blue,
urlcolor=blue,
linktocpage=true,
}

\lstdefinelanguage{pseudo}{
morekeywords=[1]{if, then, else, endif, for, while, do, loop, forall, foreach, begin, end, endforall, endforeach, endwhile, endif, endloop, return, var, pointer, to, record, break, structure, new, delete, using, struct, and},
morekeywords=[2]{unsigned, integer, int, bool, null, hword, word, void},
morekeywords=[3]{barrier_release, barrier_acquire},
keywordstyle=[3]{\bfseries\color{purple}},
sensitive=true,%
morecomment=[l]\#,%
morestring=[b]',%
}

\lstdefinelanguage{LLVM}{
morekeywords=[1]{global, define, store, load, ret, fence, return, icmp, br, label, atomic, eq, slt, alloca, sub, call, add, or, getelementptr, declare, attributes, metadata, unreachable, align, tail, common, private, unnamed_addr, constant},
morekeywords=[2]{i1, i8, i16, i32, i64, float, double, void},
morekeywords=[3]{unordered, monotonic, acquire, release, acq_rel, seq_cst},
morekeywords=[4]{@__assert_fail},
sensitive=true,%
morecomment=[l];,%
morestring=[b]"%
keywordstyle=[1]{\bfseries},
keywordstyle=[2]{\bfseries\color{blue}},
keywordstyle=[3]{\bfseries\color{purple}},
keywordstyle=[4]{\bfseries\color{red}},
commentstyle={\color{blue}},
stringstyle={\color{green!50!black}},
}

\lstdefinestyle{mine}{
morekeywords=[4]{llmc, atomic, pthread_create, pthread_join, assert, pthread_t},
morekeywords=[5]{llmc_barrier_release, llmc_barrier_acquire, llmc_barrier_seq_cst, __atomic_compare_exchange_llmc},
keywordstyle=[1]{\bfseries\color{blue}},
keywordstyle=[4]{\bfseries},
keywordstyle=[5]{\bfseries\color{purple}},
commentstyle={\color{green!50!black}},
stringstyle={\color{green!50!black}},
basicstyle={\ttfamily\scriptsize},
numbers=left,
numberstyle=\ttfamily\color{black}\bfseries,
backgroundcolor=\color{white},
mathescape=true
,literate={CASWEAK}{\casweakcode}{6}
         {CASSTRONG}{\casstrongcode}{8}
,xleftmargin=2em,framexleftmargin=1.5em}

\lstdefinestyle{pseudospecific}{style=mine
, morekeywords=[6]{hashmap,hashset}
, keywordstyle=[6]{\bfseries}
, morekeywords=[7]{construct,deconstruct,insert,get,delta,findOrPut}
, keywordstyle=[7]{\color{purple}},
literate={IL^}{\indentrulebegin}{2}
{IL|}{\indentrule}{2}
{IL>}{\indentruleend}{2}
{IL/}{\indentrulebeginend}{2},
basicstyle={\ttfamily\footnotesize},
columns=flexible,
keepspaces=true,
}

\floatstyle{ruled}
\newfloat{mylisting}{tbp}{aux}
\floatname{mylisting}{Listing}
\crefname{mylisting}{Listing}{Listings}

\DeclareCaptionSubType{mylisting}
\crefname{submylisting}{Listing}{Listings}

\newfloat{myfigure}{tbp}{aux}
\restylefloat{figure}
\floatname{myfigure}{Figure}
\crefname{myfigure}{Figure}{Figures}

\crefname{figure}{Figure}{Figures}
\crefname{table}{Table}{Tables}
\crefname{section}{Section}{Sections}

\makeatletter
\let\c@myfigure\c@figure
\makeatother

\newfloat{table}{tbp}{aux}
\floatname{table}{Table}
\newfloat{myprogram}{htbp}{aux}
\floatname{myprogram}{Program}

\newcommand{\headcell}[1]{%
\textcolor{white}{\textbf{#1}}%
}

\usepackage{pifont}

\newcommand{\sub}[1]{\ensuremath{_{\textrm{#1}}}}

\newcommand{\dmc}{{\mdseries\textsc{dmc}}\xspace}
\newcommand{\Dmc}{{\mdseries\textsc{Dmc}}\xspace}

\newcommand{\dmcmc}{{\mdseries\textsc{DMC Model Checker}}\xspace}

\newcommand{\dmcapi}{\dmc\xspace{\mdseries\textsc{api}}\xspace}

\newcommand{\nextstate}{{\mdseries\textsc{NextState}}\xspace}

\newcommand{\dtree}{{\mdseries\textsc{dtree}}\xspace}
\newcommand{\Dtree}{{\mdseries\textsc{Dtree}}\xspace}

\newcommand{\treedbs}{TreeDBS\xspace}
\newcommand{\Treedbs}{TreeDBS\xspace}

\newcommand{\treedbspad}{TreeDBS\sub{pad}\xspace}

\newcommand{\treedbsx}[1]{TreeDBS\sub{#1}\xspace}

\newcommand{\compacttree}{Compact Tree\xspace}
\newcommand{\Compacttree}{Compact Tree\xspace}

\newcommand{\promela}{{\mdseries\textsc{Promela}}\xspace}

\makeatletter
\newcommand{\provides}[1]{%
\expandafter\def\csname{textprovide#1}\endcsname{1}%
\hypertarget{#1}{}%
\marginpar{{\scriptsize\color{green!60!black}+#1}}%
}
\WithSuffix\newcommand\provides*[1]{%
\expandafter\def\csname{textprovide#1}\endcsname{1}%
\hypertarget{#1}{}%
\marginpar{{\scriptsize\color{green!60!black}+#1}}%
}
\makeatother

\newcommand{\needs}[1]{%
\ifcsname{textprovide#1}\endcsname%
#1%
\marginpar{{\scriptsize\hyperlink{#1}{\textcolor{blue}{>#1}}}}%
\else%
\textcolor{red}{[#1]}%
\marginpar{{\scriptsize{\textcolor{red}{>#1}}}}%
\fi%
}

\WithSuffix\newcommand\needs*[1]{%
\ifcsname{textprovide#1}\endcsname%
\marginpar{{\scriptsize\hyperlink{#1}{\textcolor{blue}{>#1}}}}%
\else%
\textcolor{red}{[#1]}%
\marginpar{{\scriptsize{\textcolor{red}{>#1}}}}%
\fi%
}

\usepackage{mathtools}

\definecolor{indentlinecolor}{HTML}{CCCCCC}
\newcommand{\indentrulebegin}{\textcolor{indentlinecolor}{\rlap{\smash{\rule[-.35em]{.7pt}{1.3em}}}}}
\newcommand{\indentrule}{\textcolor{indentlinecolor}{\rlap{\smash{\rule[-.35em]{.7pt}{1.45em}}}}}
\newcommand{\indentruleend}{\textcolor{indentlinecolor}{\rlap{\smash{\rule[-.35em]{.7pt}{1.45em}\rule[-.35em]{.5em}{.8pt}}}}}
\newcommand{\indentrulebeginend}{\textcolor{indentlinecolor}{\rlap{\smash{\rule[-.35em]{.7pt}{1.3em}\rule[-.35em]{.5em}{.8pt}}}}}

\newcolumntype{R}[1]{>{\raggedleft\let\newline\\\arraybackslash\hspace{0pt}}m{#1}}

{\color{blue}}%
{}

\usepackage[clock]{ifsym}

\newcommand{\resulttimeout}[3]{%
\ifthenelse{\equal{#1}{0}}{%
\raisebox{-1.6pt}{\showclock{#2}{#3}}%
} {%
\raisebox{-1.6pt}{\showclock{#2}{#3}\textsubscript{+#1}}%
}%
}



\graphicspath{{../../images/hashtable/}{./images/}{./images/histograms/}{./images/beemresults/}{../../experiments/llmc/}{../../experiments/llmc/symbols/}{../../../experiments/llmc/symbols/}}

\definecolor{colone}       {HTML}{8DE9FF}
\definecolor{coltwo}       {HTML}{9D9BFF}
\definecolor{colthree}     {HTML}{A6FF8A}
\definecolor{colfour}      {HTML}{FFC48A}
\definecolor{colonedark}   {HTML}{78A1AB}
\definecolor{coltwodark}   {HTML}{7F7EAB}
\definecolor{colthreedark} {HTML}{83AB76}
\definecolor{colfourdark}  {HTML}{AB9076}
\definecolor{colonesat}    {HTML}{1A7489}
\definecolor{coltwosat}    {HTML}{1E1C89}
\definecolor{colthreesat}  {HTML}{348919}
\definecolor{colfoursat}   {HTML}{895119}
\definecolor{colonelight}  {HTML}{B3F0FF}
\definecolor{coltwolight}  {HTML}{BDBCFF}
\definecolor{colthreelight}{HTML}{C3FFB0}
\definecolor{colfourlight} {HTML}{FFD7B0}

\definecolor{colabcd}        {HTML}{C3FFB0}
\definecolor{colabcdef}      {HTML}{9D9BFF}
\definecolor{colabghef}      {HTML}{FFC48A}
\definecolor{colabcdefghij}  {HTML}{B3F0FF}
\definecolor{colabcdefghijk} {HTML}{FFC48A}
\definecolor{colabcdefghijkl}{HTML}{C3FFB0}

\definecolor{proccolor}    {HTML}{B3F0FF}
\definecolor{memcolor}     {HTML}{BDBCFF}
\definecolor{bufcolor}     {HTML}{C3FFB0}
\definecolor{thingcolor}   {HTML}{1A7489}
\definecolor{thingcolortwo}{HTML}{89d8e8}
\definecolor{redcolor}     {HTML}{FFCFD6}
\definecolor{reddark}      {HTML}{891929}
\definecolor{toolcolor}    {RGB} {230,230,230}

\newcommand{\collapse}{{\scshape{Collapse}}\xspace}

\newcommand{\divine}{\mbox{\mdseries\textsc{divine}}\xspace}
\newcommand{\Divine}{\mbox{\mdseries\textsc{Divine}}\xspace}

\newcommand*\vectorbox[2]{%
\setlength{\fboxsep}{1pt}%
\fcolorbox{black}{#1}{\protect\vphantom{X}\small\texttt{#2}}%
}

\usepackage{tcolorbox}

\newcommand*\indexbox[2]{%
\tcbox[nobeforeafter,tcbox raise base,boxrule=0.4pt,top=0mm,bottom=0mm,right=0mm,left=0mm,arc=1pt,boxsep=1pt,colframe=black,colback=#1]{%
\protect\vphantom{X}\small\texttt{#2}
}%
}

\newcommand*\indexlengthbox[3]{%
\tcbox[nobeforeafter,tcbox raise base,boxrule=0.4pt,top=0mm,bottom=0mm,right=0mm,left=0mm,arc=1pt,boxsep=1pt,colframe=black,colback=#1]{%
\protect\vphantom{X}\small\texttt{#2}
}${}_{\mbox{#3}}$%
}

\usepackage[kerning=true,stretch=10,shrink=30]{microtype}

\crefname{figure}{fig.}{Figures}
\Crefname{figure}{Figure}{Figures}

\begin{document}
\title{\doctitle\\\docsubject}
\titlerunning{\doctitle}
\author{\docauthor}
\authorrunning{\docauthorrunning}
\institute{\docaff \\
\email{\docemail}}
\maketitle              
\begin{abstract}

High-performance software typically uses dynamic memory allocations and multi-threading to leverage multi-core CPUs. Model checking such software not only has to deal with state space explosion, but also with variable-length states due to dynamic allocations. Moreover, changes between states are typically small, calling for incremental updates. Many model checkers, although efficiently dealing with the latter, only support fixed-length state vectors. In this paper, we introduce \dtree, a concurrent compression tree data structure that compactly stores variable-length states while allowing partial state reconstruction and incremental updates without reconstructing states. We implemented \dtree in the \dmc multi-core model checker. We show that, for models with states of varying length, \dtree is up to 2.9 times faster and uses on average 29\% less memory than state-of-the-art tools.

\end{abstract}

\setlength{\floatsep}{0pt}%
\setlength{\textfloatsep}{0pt}%
\setlength{\intextsep}{0pt}%

\section{Introduction}
\label{sec:intro}

High-performance concurrent software is complex to write and even harder to reason about.
The more threads run in parallel, the more different interleavings are possible, easily causing billions (or trillions) of reachable program states.
Programmers want to make sure that they are valid, i.e.\ none represent an error due to stack overflows, race conditions, buffer overruns, null pointer dereferences, or other erroneous operations.
One way to verify such properties is to \emph{model check}~\cite{DBLP:reference/mc/ClarkeHV18} the program, during which the entire state space is explored.
We can remember the visited states to avoid visiting them multiple times and thus avoid doing the same work redundantly.
A complication when model checking software is dealing with variable-length states, due to dynamic memory allocations, e.g. heap memory.
However, changes between states are typically small. Model checkers can take advantage of this for efficiency.
Thus, model checking software has four major requirements:
\begin{wrapfigure}[8]{r}{.25\columnwidth}%
\centering%
\def\svgwidth{.375\textwidth}%
\scalebox{.6}{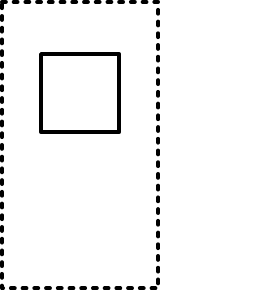}%
\label{fig:intro:dmc}%
\end{wrapfigure}
\textbf{R1}) efficiently store billions or trillions of states;
\textbf{R2}) efficiently calculate successor states;
\textbf{R3}) efficiently determine whether or not a state has been visited;
\textbf{R4}) support states of varying length, due to dynamic memory.
These requirements have implications on how states are stored, but also on how next states are determined and communicated 
between model, model checker and state storage, in a modularized implementation as shown on the right.

\subsection{Related work}
\label{sec:intro:relatedwork}

LTSmin~\cite{10.1007/978-3-662-46681-0_61}, together with its compression trees \treedbs~\cite{DBLP:conf/spin/LaarmanP011} and \compacttree~\cite{DBLP:journals/isse/Laarman19} is one example on how to approach these requirements in a modular implementation. LTSmin is a model checker with multi-core~\cite{ltsmin-mc:nmf2011} (up to 64 threads), distributed~\cite{eemcs18152} and symbolic implementations~\cite{DBLP:conf/ictac/BlomP08}.
These implementations share a common interface: PINS, the Partitioned Next-State Interface, their API between search core and model.
PINS does not lend itself well for software, because it only support fixed-length states. Moreover, it requires the complete state to be available when the model is asked for successor states.
Because these states are stored in \treedbs or \compacttree, these states need to be reconstructed first, which comes at a price. Communicating a next state is also done using a complete state. LTSmin can use a projection~\cite{10.1007/978-3-662-46681-0_61} to detect the changed parts to minimize copying.
However, this projection requires static information on which parts of the state are touched, which is not guaranteed available in software due to the dynamic nature of heap memory and spawning threads dynamically.
LTSmin does have a software front-end~\cite{essay65059}, but since LTSmin only supports fixed-length states, that front-end uses a sub-optimal uncompressed chunk table to model heap memory, which also lacks projection.

\treedbs~\cite{DBLP:conf/spin/LaarmanP011} is a thread-safe state storage for fixed-length states.
It uses a binary compression tree, based on work by Blom et al.~\cite{DBLP:journals/logcom/BlomLP011}, in order to identify common sub-vectors.
Laarman et al. show it to be efficient (R1, R2 and R3), but it does not support states of varying length (R4).
\Compacttree~\cite{DBLP:journals/isse/Laarman19} is an evolution of \treedbs, where a Cleary table~\cite{10.1109/TC.1984.1676499} is used for its root set. We will explain \treedbs and \compacttree in more detail in \cref{sec:background:treedbs}.

SPIN version 6~\cite{588521} is an explicit-state model checker for \promela models. It supports a dynamic number of \promela processes by having a variable length root state. Moreover, it has a state compression method \collapse~\cite{Holzmann97statecompression}: instead of fully storing all combinations of \promela process states in the root state, the states of the processes themselves are stored and mapped to a unique ID, and those IDs are stored in the root state instead.
Thus, a state is a two-level tree of states: a root state and the process states. This is similar to using the chunk table of LTSmin.
SPIN supports a number of search algorithms, among which is parallel BFS up to 63 threads.

\Divine version 4~\cite{DBLP:journals/jss/RockaiSCB18} is an explicit-state model checker for LLVM IR assembly code~\cite{Lattner:MSThesis02}. It uses a graph to model the stack and heap. Each node in this graph represents an allocated section of memory, for example space allocated for a \code{struct}. Edges on this graph represent reachability, e.g. a pointer in a \code{struct} to another \code{struct}.

Other, more remote approaches to state compression, are Binary Decision Diagrams~\cite{DBLP:journals/tc/Bryant86,DBLP:conf/dac/BurchCMD90} and PTries~\cite{DBLP:conf/ictac/JensenLS17}.
BDDs compress by sharing prefixes and suffixes of (Boolean) state vectors, and are used in symbolic model checking; in their standard form, they are restricted to static vector lengths.
PTries compress by only sharing prefixes of subvectors; they do support dynamic state lengths natively.
Laarman compares the compression and performance of BDDs and PTries to \compacttree~\cite{DBLP:journals/isse/Laarman19}.

\subsection{Contributions}

In this paper we introduce \dtree, a state storage data structure that satisfies R1--R4: it is concurrent, compresses states, can handle variable-length states and provides partial reconstruction of a stored state and incremental updates to states \emph{without the need for reconstructing the entire state}.
We implemented \dtree as part of the \dmcmc.

To showcase the potential of \dtree, we compare it to other state storage components using models with variable-length states. In the case of \treedbs, we \emph{pad} all vectors to the length of the largest vector (manually determined). Here, \dtree is able to perform up to 2.9 times faster than \treedbs using 29\% less memory on average, without the a priori need to manually determine the largest vector.
The advantage of \dtree increases as the difference in length between states increases.
To evaluate how much efficiency is lost to gain variable-length support, we compare \dmc using \dtree to LTSmin and SPIN using models from the BEEM database~\cite{10.1007/978-3-540-73370-6_17} with fixed-length states. 
In this setting, \dtree is actually marginally faster than \treedbs, but uses 30\% more memory.
\Compacttree compresses fixed-length states 2.3x more than \dtree.
Compared to SPIN using \collapse, \dtree is 8.1x faster and uses 6.1x less memory.

\setlength{\floatsep}{8pt}%
\setlength{\textfloatsep}{10pt}%
\setlength{\intextsep}{10pt}%

\section{Tree compression}

\begin{figure}[!b]
\centering%
\def\svgwidth{.8\textwidth}
\scalebox{1.0}{\ttfamily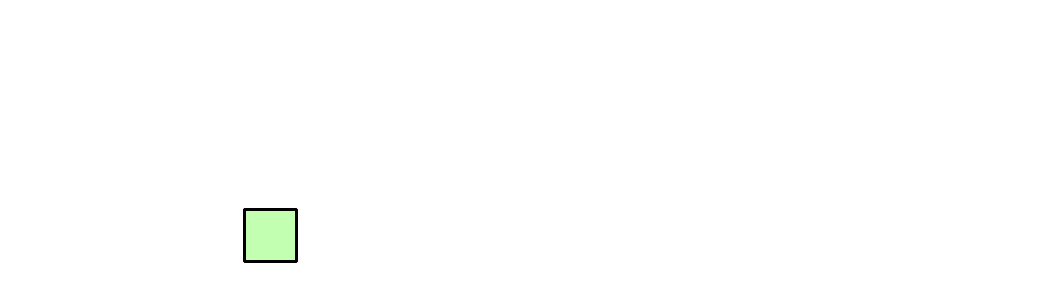}
\caption{The layout of the indexed hash set after inserting \vectorbox{colabcd}{abcd} and \vectorbox{colabcdef}{abcdef}.
}%
\label{fig:dtree:abcd_map}%
\end{figure}

We now explain \treedbs and \compacttree, which inspired \dtree.
States are vectors of \emph{state slots}; a variable in a model occupies one or more state slots.
For the sake of simplicity, let us assume these state slots are 32-bit.
The principle behind the compression is to insert pairs of state slots, thus 64-bit values, into an \emph{indexed hash set} with a 32-bit index.
In an indexed hash set, the index is determined by \emph{hashing} the value.
A pair of 32-bit indices forms another 64-bit value, which in turn can be put into the set.
This process continues, creating an ordered tree of \emph{nodes}.
This is illustrated in \cref{fig:dtree:abcd_map}: the vectors \vectorbox{colabcd}{abcd} and \vectorbox{colabcdef}{abcdef} share a common sub-vector \vectorbox{white}{abcd} and index \vectorbox{white}{3} maps to this sub-vector.
This also illustrates the compression: common nodes are only inserted once.
The higher up the common node is in the tree, the larger the common sub-vector it maps to, the more memory is saved.
Which common sub-vectors can be identified and reused, depends on the structure of the binary tree: while \vectorbox{white}{abcd} can be identified, \vectorbox{white}{cdef} cannot.
If we were to add the vector \vectorbox{white}{cdef}, we would add the node \vectorbox{white}{24}. Similarly, we would add the node \vectorbox{white}{42} for \vectorbox{white}{efcd} and the node \vectorbox{white}{33} for \vectorbox{white}{abcdabcd}.

Note that the rounded boxes \indexbox{colabcd}{3} and \indexbox{colabcdef}{5} are the indices to the root nodes of the trees mapping to the states. However, an index is not enough to identify a state, as index \indexbox{white}{3} could map to \vectorbox{white}{12} as well. Indeed, if we would have added~the~state~\vectorbox{colfour}{12} (with index \indexbox{colfour}{3}), this would be completely correct, thus we need more information other than \indexbox{white}{3} to distinguish \indexbox{colabcd}{3} from \indexbox{colfour}{3}. This gives rise to two issues: uniquely identifying a state and determining whether or not a state has already been added.

\subsection{\treedbs and its implementation}
\label{sec:background:treedbs}

\begin{figure}[!b]
\begin{subfigure}{.48\textwidth}
\centering%
\def\svgwidth{1.1\columnwidth}
\scalebox{.9}{\ttfamily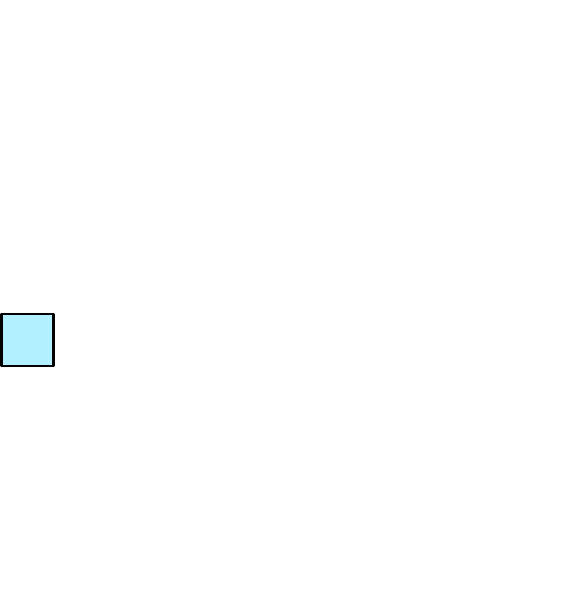}
\caption{Laarman et al.'s compression tree schema as in \emph{their paper}.
The top state adds 7 nodes, bringing the total to 16.
}%
\label{fig:dtree:poweroftwodiff:laarmanpaper}%
\end{subfigure}\hfill%
\begin{subfigure}{.48\textwidth}
\centering%
\def\svgwidth{1.1\columnwidth}
\scalebox{.9}{\ttfamily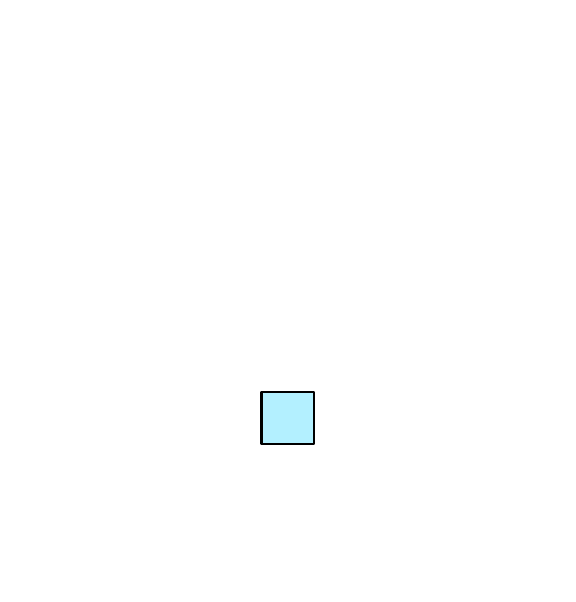}
\caption{Laarman et al.'s compression tree schema as in \emph{their implementation}.
The top state adds 10 nodes; 19 nodes in total.
}%
\label{fig:dtree:poweroftwodiff:laarmanimpl}%
\end{subfigure}
\caption{An illustration of the difference between the \emph{paper} and \emph{implementation} of the compression tree of Laarman et al. Given a 40 byte state \vectorbox{colabcdefghij}{abcdefghij} (bottom tree), the next state \vectorbox{colabcdefghijk}{abcdefghijk} (top tree) appends \vectorbox{colabcdefghijk}{k}, 4 bytes.}
\label{fig:dtree:poweroftwodifflaarman}%
\end{figure}

\Treedbs solves the first issue by only supporting fixed-length states and having the length of the state dictate the shape of the tree. Thus, all trees are isomorphic and an index uniquely identifies a state. However, the implementation~\cite{ltsmingit} of how the length dictates the shape of the tree differs from the how it was originally presented~\cite{DBLP:conf/spin/LaarmanP011}.
We show both versions in \cref{fig:dtree:poweroftwodifflaarman}, ignoring \vectorbox{colabcdefghijk}{\protect\phantom{X}} (top) and only focusing on \vectorbox{colabcdefghij}{\protect\phantom{X}} (bottom) for now.
The paper version recursively divides the vector in two, associating one half with the left child of a node and the other half with the right child (\cref{fig:dtree:poweroftwodiff:laarmanpaper}). For efficiency, the implementation version represents the tree as an array, where a node at index $n$ has its children at indices $2n$ and $2n+1$ (\cref{fig:dtree:poweroftwodiff:laarmanimpl}). Inserting the nodes of a vector then becomes a simple for-loop since children are always adjacent. Both versions produce a balanced tree, but their balancing is different, influencing locality. For example, the paper version has a single pair of adjacent state slots (\vectorbox{colabcdefghij}{ef}) that requires traversing both the left and right side of the tree, whereas the implementation version has two (\vectorbox{colabcdefghij}{bc} and \vectorbox{colabcdefghij}{fg}). The importance of this will become apparent in \cref{sec:dtree:incupdates}.

To determine whether a state has been added already, the paper version of \treedbs uses a single indexed hash set and a root-bit in the nodes to indicate whether a node is a root node or a tree node. This distinction is needed because otherwise a leaf node could be mistaken for a root node and it could be erroneously concluded a state was already added.
The current implementation removes the root-bit in favour of using two hash sets: a root hash set that stores the root nodes and a hash set that stores all other nodes.
Not needing the root-bit, 32-bit indices can be used instead of just 31-bit indices and this lifts the 31-bit restriction on the data in state slots as well.
In addition, it allows the root set to be larger than $2^{32}$, since its indices need not fit in 32 bits. The data set is still limited to 32-bit indices, because paired they are limited to 64 bits.
However, one specific 64-bit value in the hash map is still reserved to indicate the \emph{empty value}. This value is needed to distinguish between an empty bucket and a used bucket. In the case of \treedbs, this is the value $-1$, which means a model cannot pair two state slots that have a $-1$ value, because together they would form a 64-bit $-1$ value. As we will see in the experiments, this does happen.

\paragraph{\Compacttree}\cite{DBLP:journals/isse/Laarman19} is the same as \treedbs, with the root set replaced by a Cleary Table~\cite{10.1109/TC.1984.1676499}. \Treedbs supports root set indices of fewer than 64 bits, but \compacttree can only return a 64-bit root index, due to the nature of the Cleary Table. Furthermore, the paper on \compacttree shows the same manner of determining the shape of the tree as the \treedbs paper, but the implementation is shared with \treedbs, with a flag to enable the use of a Cleary Table. This means Compact Tree suffers from the same alignment issues as \treedbs.

\section{Dtree}

Like \treedbs, we base our compression tree on the compression tree of Blom et al.~\cite{DBLP:journals/logcom/BlomLP011} and use two indexed hash sets. We diverge on a number of crucial points from \treedbs in order to meet the requirements mentioned in \cref{sec:intro}. Since \treedbs supports only fixed-length vectors, there is no need to remember the length of each individual vector, but we do (R4).
We also introduce a different compression tree structure that benefits dynamic memory allocation (R4).
Furthermore, we extend the capabilities of the compression tree to store and reconstruct only parts of states in order to improve calculating next states (R2). Lastly, our approach supports any data, including $-1$ values in adjacent state slots.

First, we show why we need a different tree structure.
Let us consider the scenario where a model has allocated 40 bytes (10 state slots) and in the next state the model wishes to allocate 4 more bytes. The model does this by growing the state by one state slot. \Cref{fig:dtree:poweroftwodifflaarman} shows how this scenario is supported by the two versions of \treedbs.
The implementation version starts forming 64-bit sections starting at the end, so a change of 4 bytes (32 bits) to the size of the state will unalign the tree nodes compared to the previous state. This results in storing the same information twice. This effect is not limited to the last level of the tree: a change of 8 bytes in the size of the state results in the second to last level of the tree to become unaligned. A change of 16 bytes would unalign the third to last level, etc.
\Cref{fig:dtree:poweroftwodiff} shows two improvements to these structures. \Cref{fig:dtree:poweroftwodiff:laarmanimplbackwards} describes the structure of the implementation version of \treedbs, but backwards in an attempt to limit the alignment issue. This works for the leaves, but one level higher in the tree we face the same issue: instead of \vectorbox{colabcdefghij}{12}, \vectorbox{colabcdefghij}{34} we get \vectorbox{colabcdefghijk}{23}, \vectorbox{colabcdefghijk}{45}. Moreover, this approach still has two pairs of adjacent state slots that require traversing both the left and right side of the tree (\vectorbox{colabcdefghij}{de} and \vectorbox{colabcdefghij}{hi}).

\subsection{A Chain of Perfectly Balanced Binary Trees}
\label{sec:dtree:chain}

\begin{figure}[!b]
\begin{subfigure}{.48\textwidth}
\centering%
\def\svgwidth{1.1\columnwidth}
\scalebox{.9}{\ttfamily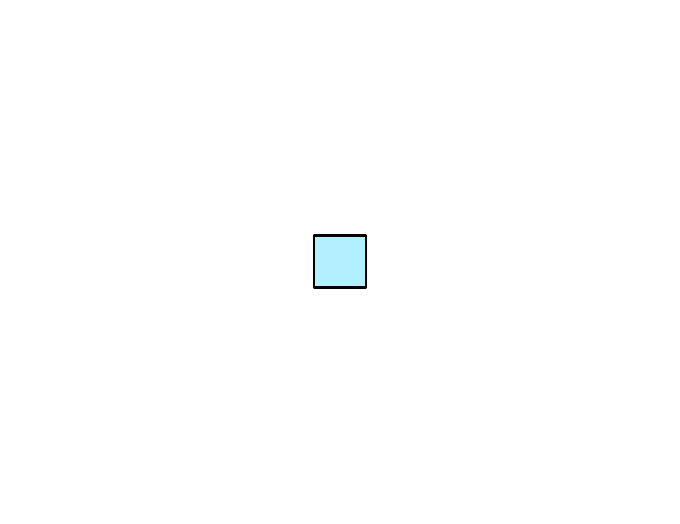}
\caption{Laarman et al.'s compression tree schema as in \emph{their implementation}, but \emph{left-to-right} instead of right-to-left.
The next state adds 5 nodes; 14 nodes in total.
}%
\label{fig:dtree:poweroftwodiff:laarmanimplbackwards}%
\end{subfigure}\hfill%
\begin{subfigure}{.48\textwidth}%
\vspace{1.57cm}%
\centering%
\def\svgwidth{1.1\columnwidth}%
\scalebox{.9}{\ttfamily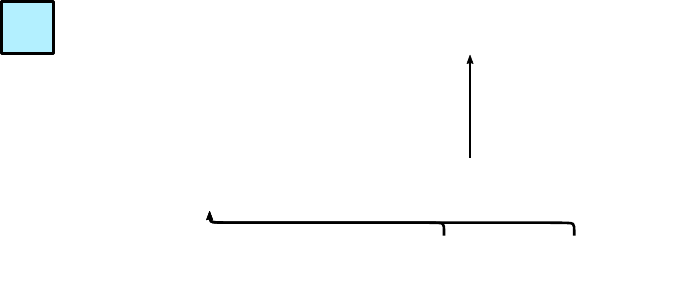}%
\caption{The two states using our chain of perfectly balanced binary trees schema.
The next state adds 2 nodes, bringing the total to 11. Additionally, it shows a third state \vectorbox{colabcdefghijkl}{\protect\phantom{X}} when \vectorbox{colabcdefghijkl}{l} is appended.
}%
\label{fig:dtree:poweroftwodiff:ours}%
\end{subfigure}
\caption{Like \cref{fig:dtree:poweroftwodifflaarman}, given a 40 byte state \vectorbox{colabcdefghij}{\protect\phantom{X}}, the next state \vectorbox{colabcdefghijk}{\protect\phantom{X}} appends \vectorbox{colabcdefghijk}{k}, 4 bytes.}
\label{fig:dtree:poweroftwodiff}%
\end{figure}

\Cref{fig:dtree:poweroftwodiff:ours} shows our approach, named \emph{the chain of perfectly balanced binary trees}. This approach does not suffer from the issues of the other tree shapes as \cref{fig:dtree:poweroftwodiff:ours} clearly demonstrates. In a chain, the length of the sub-vector the left child of a node leads to is always $\mbox{lpst}(\mbox{len}(V))$, where $\mbox{lpst}(x)$ is the \textbf{l}argest \textbf{p}ower-of-two \textbf{s}maller \textbf{t}han $x$, $\mbox{len}(V)$ is the length of the vector $V$ and $V$ is the vector that the node itself leads to.
Thus, a left child is always perfectly balanced. The right part maps to the remainder of the vector.
For example, the state \vectorbox{colabcdefghijk}{abcdefghijk} of length 11 in \cref{fig:dtree:poweroftwodiff:ours}, is shaped by perfectly balanced trees mapping to 8, 2 and 1 state slots; the state \vectorbox{colabcdefghijkl}{abcdefghijkl} is shaped by trees mapping to 8 and 4 state slots. This can result in a less balanced tree than \treedbs. For example, for the state \vectorbox{colabcdefghij}{abcdefghij}, \treedbs (\cref{fig:dtree:poweroftwodiff:laarmanimpl}) has 2 nodes on the lowest level (\vectorbox{colabcdefghij}{gh}, \vectorbox{colabcdefghij}{ij}) and 3 on the second-to-lowest level (\vectorbox{colabcdefghij}{ab}, \vectorbox{colabcdefghij}{cd}, \vectorbox{colabcdefghij}{ef}). \Dtree (\cref{fig:dtree:poweroftwodiff:ours}) on the other hand has 4 on the lowest level (\vectorbox{colabcdefghij}{ab}, \vectorbox{colabcdefghij}{cd}, \vectorbox{colabcdefghij}{ef}, \vectorbox{colabcdefghij}{gh}) and one on the second-to-highest level (\vectorbox{colabcdefghij}{ij}). This means changes to \vectorbox{colabcdefghij}{ab}, \vectorbox{colabcdefghij}{cd} or \vectorbox{colabcdefghij}{ef} require \dtree an \emph{extra} node and changes to \vectorbox{colabcdefghij}{ij} require two \emph{fewer} nodes, compared to \treedbs.
We will investigate the impact of the difference in \cref{sec:experiments}.
\Cref{fig:dtree:poweroftwodiff:ours} illustrates that this approach does lend itself to appending to states.
This targets a combination of R1 and R4 from \cref{sec:intro}: to more efficiently handle dynamic memory allocation.
It also has a single pair of adjacent state slots requiring traversing both sides of the tree (\vectorbox{colabcdefghij}{hi}).

\subsection{Incremental updates}
\label{sec:dtree:incupdates}

The importance of minimizing such adjacent pairs becomes apparent when adding support for incremental updates. With this, a model conveys to the model checker only what \emph{changed} compared to a previous state, in lieu of a complete state.
\treedbs supports incremental updates by \emph{detecting} unchanged parts, when inserting a complete state, but it does not support simply applying an incremental update to a state. We describe the difference with an example. Considering the state \vectorbox{colabcdef}{abcdef}, the model determines that in the next state \vectorbox{colabcdef}{cd} changes to \vectorbox{colabghef}{gh}. We have two options of communicating this to the model checker:
\begin{enumerate}[noitemsep]
\item the model can communicate the entire new state \vectorbox{colabghef}{abghef}; or
\item the model can communicate only the change of \vectorbox{colabghef}{gh} at offset 2, denoted \vectorbox{colabghef}{gh}@2.
\end{enumerate}

\noindent
For small states the difference in performance is likely negligible. However, since we want to model check software, our states can be quite large and thus this could save a significant amount of copying.

When communicating only changes, the model checker also needs to know which state the change needs to be applied to. \vectorbox{colabcdef}{34} (as per \cref{fig:dtree:abcd_map}) alone does not identify \vectorbox{colabcdef}{abcdef}: we need the length of the vector as well.

We considered three ways to solve this:
\setlist{nolistsep}
\begin{enumerate}[noitemsep]
\item Always remember the length alongside the index. This has the downside that high-performance low-level atomic instructions such as \code{compare-and-swap} cannot be used. 128-bit atomic instructions are more expensive~\cite{intel64manual}.
\item Add another node to the tree, containing the length and index to the remainder of the tree. Thus, the index to that node would uniquely identify a state. This adds a level to the tree, which can negatively influence performance.
\item Remember index and length, but limit their combination to 64 bits. For example, 40 bits for the index and 24 for the length. This has the disadvantage of only supporting a trillion states of $2^{24}-1$ state slots each. At the current technological age, both seem upperbounds on what we need for the foreseeable future, so this should suffice.
\end{enumerate}

\begin{figure}[!b]
\begin{subfigure}{.5\textwidth}
\centering%
\def\svgwidth{1.25\columnwidth}
\scalebox{.7}{\ttfamily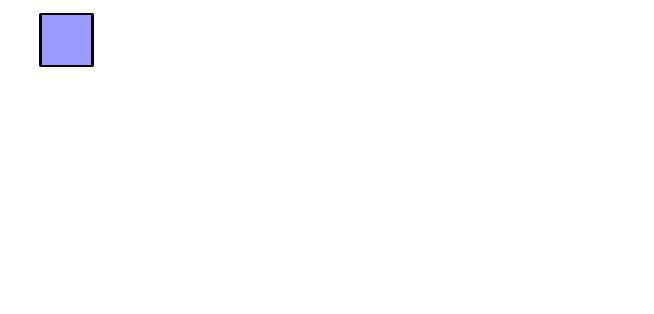}
\caption{\code{delta()} is called recursively}%
\label{fig:dtree:abghef:1}
\end{subfigure}
\begin{subfigure}{.5\textwidth}
\centering%
\def\svgwidth{1.25\columnwidth}
\scalebox{.7}{\ttfamily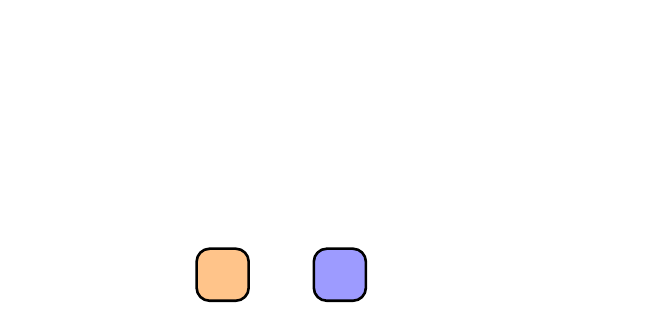}
\caption{Applied delta \vectorbox{colabghef}{gh}@2 to \vectorbox{colabcdef}{abcdef}}%
\label{fig:dtree:abcdefgh_rootmap}%
\end{subfigure}
\caption{The delta-vector \vectorbox{colabghef}{gh} is applied at offset 2 to \vectorbox{colabcdef}{abcdef} from \cref{fig:dtree:abcd_map}.}
\label{fig:dtree:abghef}%
\end{figure}
\noindent
We chose the third alternative. We call this combination the \code{StateID} and we write this for example as \indexlengthbox{colabcdef}{5}{6}, which uniquely identifies \vectorbox{colabcdef}{abcdef}: 5 is the index, 6 is the length of the state.
Applying a delta (incremental update) such as \vectorbox{colabghef}{gh}@2 to \indexlengthbox{colabcdef}{5}{6} is described in \cref{fig:dtree:abghef}.
First, in \cref{fig:dtree:abghef:1}, we recursively traverse the tree to find the leaves that need change. At each node that leads to vector $V$, we check which of the children are affected. From \cref{sec:dtree:chain} follows that the left child is affected iff the offset is smaller than $\mbox{lpst}(\mbox{len}(V))$. The right child is affected iff the offset plus the length of the delta is greater than $\mbox{lpst}(\mbox{len}(V))$. Thus, the tree is traversed as follows:
\begin{itemize}
\item[] \vectorbox{white}{gh}@2 to \indexlengthbox{colabcdef}{5}{6} only affects the left child, since $2<\mbox{lpst}(6)$ and $2+2 \not>\mbox{lpst}(6)$
\begin{itemize}
\item[] \vectorbox{white}{gh}@2 to \indexlengthbox{colabcdef}{3}{4} only affects the left child, since $2\not<\mbox{lpst}(4)$ and $2+2 >\mbox{lpst}(4)$
\begin{itemize}
\item[] \vectorbox{white}{gh}@0 to \indexlengthbox{colabcdef}{2}{2} completely replaces the leaf node.
\end{itemize}
\end{itemize}
\end{itemize}
\noindent
At this point we can traverse back, inserting new nodes along the path. First we insert the new node \vectorbox{colabghef}{gh} at index \indexbox{colabghef}{6} in the data set, then we use that index to create a new node \vectorbox{colabghef}{16}, where we copy the index of the left child from \vectorbox{colabcdef}{12}. Finally, we insert the root node \vectorbox{colabghef}{74}, providing \code{StateID} \indexlengthbox{colabghef}{8}{6} (\cref{fig:dtree:abcdefgh_rootmap}).
A delta beyond the length of a state is also supported. For example, we could have applied \vectorbox{white}{gh}@5 to \indexlengthbox{colabcdef}{5}{6}, resulting in \vectorbox{white}{abcdegh}. When applying \vectorbox{white}{gh}@8 to \indexlengthbox{colabcdef}{5}{6} it would yield \vectorbox{white}{abcdef00gh}.

\subsection{Partial reconstruction of states}
\label{sec:dtree:reconstruct}

In a similar fashion to incremental updates, we can reconstruct (parts of) states. We recursively traverse the tree of nodes until we reach the correct leaf nodes. Then, we copy the contents of these leaf nodes into a buffer, yielding (a part of) a state.

\subsection{Determining a state is new}

When creating a new state, either by inserting a complete new one or by incremental update, the model checker needs to know if the state has already been visited or not. This is accomplished by use of the root set. A state has already been inserted into \dtree iff there is a corresponding root node. As an example, let us insert \vectorbox{colabcdef}{abcdef} into the \dtree of \cref{fig:dtree:abcdefgh_rootmap}, which already exists. Recursively, we traverse the nodes. At the leaves, we conclude \vectorbox{white}{ab}, \vectorbox{white}{cd} and \vectorbox{white}{ef} have already been inserted.
The tree node \vectorbox{white}{12} also has already been inserted. Then we conclude that the root node \vectorbox{white}{34} has been inserted as well.
Now let us insert \vectorbox{colabcd}{abcd}. Again, we conclude the leave nodes \vectorbox{white}{ab} and \vectorbox{white}{cd} are already in the tree. Even though \vectorbox{white}{12} is already in the data set, it is not yet in the root set and thus we correctly conclude it is a new state. Additionally, \dtree supports inserting states/vectors that are inserted purely in the data set and not its root node in the root set. This can be useful to compress data that is not a state in the model as we will see later in \cref{sec:dmc:example}. Thus, \dtree provides the following API:
{\lstset{language=pseudo,style=pseudospecific,numbers=none,aboveskip=2pt,belowskip=0pt}
\begin{lstlisting}
StateID insert(Slot[] V, bool root)
StateID delta(StateID s, int offset, Slot[] D, bool root)
Slot[]  get(StateID s, bool root)
Slot[]  get(StateID s, int offset, int length, bool root)
\end{lstlisting}}
\noindent
Here, \code{root} indicates if the root node should indeed be placed in the root set. Note that this is also needed for \code{get()} because we need to look for the root node in the right set. The interfaces \code{insert()} and \code{delta()} insert a new state (\cref{sec:dtree:incupdates}). The two \code{get()} interfaces allow to obtain (parts of) the state (\cref{sec:dtree:reconstruct}).

\setlength{\floatsep}{1pt}%
\setlength{\textfloatsep}{1pt}%
\setlength{\intextsep}{1pt}%

\clearpage
\section{DMC Model Checker}

\begin{mylisting}[!b]
\centering%
\begin{subfigure}{.48\textwidth}
\centering%
{\lstset{language=pseudo,style=pseudospecific,xleftmargin=1.5em,framexleftmargin=0em,numbersep=4pt}
\begin{lstlisting}
Q = {};
M.initialState();
while(!Q.isEmpty())
IL^StateID s = Q.pop();
IL>M.nextStates(s);
\end{lstlisting}}
\caption{A basic search core that simply requests next states from a model \code{M} until all have been visited. For brevity, the on-the-fly checking of properties is left out.}%
\label{lst:dmc:core}%
\end{subfigure}\hspace{10pt}%
\begin{subfigure}{.44\textwidth}
\centering%
{\lstset{language=pseudo,style=pseudospecific,xleftmargin=1em,framexleftmargin=0em,numbersep=4pt}
\begin{lstlisting}
void initialState()
IL^MC.insert({0,0,0,0}, true);
IL># ^ implicitly pushes to Q
void nextStates(StateID s)
IL|for i in {0,1,2,3}
IL|IL^Slot v = MC.get(s, i, 1);
IL|IL|v = (v + 1) % 10;
IL>IL>MC.delta(s, i, {v}, true);
\end{lstlisting}}
\caption{A model with four counters.
}%
\label{lst:dmc:countermodel}%
\end{subfigure}
\caption{A search core and a simple model with a state space size of $10^4$.}%
\label{lst:dmc:coreandmodel}%
\end{mylisting}
\begin{mylisting}[!b]
\centering%
{\lstset{language=pseudo,style=pseudospecific,xleftmargin=1.5em,framexleftmargin=0em,numbersep=4pt}
\begin{lstlisting}
StateID delta(StateID s, int offset, Slot[] D, bool root)
IL^{StateID s, bool isNew} = DTREE.delta(s, offset, D, root);
IL|if(isNew and root) Q.push(s);
IL>return s;
\end{lstlisting}}
\caption{An implementation of the \code{delta()} interface of the \dmcapi.}%
\label{lst:dmc:deltaimpl}%
\end{mylisting}

\begin{wrapfigure}[10]{r}{.3\columnwidth}%
\centering%
\def\svgwidth{.42\textwidth}%
\scalebox{.7}{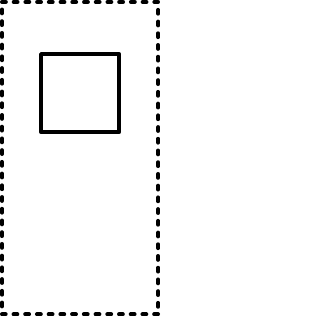}%
\caption{The modular DMC Model Checker.}%
\label{fig:dmc:modular}%
\end{wrapfigure}

We now expand on the concept of a modular model checker as described in \cref{sec:intro}. A model \code{M} implements the \nextstate (NS) API:

{\lstset{language=pseudo,style=pseudospecific,numbers=none,aboveskip=2pt,belowskip=0pt}
\begin{lstlisting}
void initialState()
void nextStates(StateID s)
\end{lstlisting}}

\noindent
Here, \code{initialState()} sets up the initial state and \code{nextStates()} computes the next states of the specified state \code{s}. Communicating (``uploading'') states to the model checker is done using the \dtree API, which is exposed to the model via the search core. This flow is illustrated in \cref{fig:dmc:modular}.
\setlength{\floatsep}{8pt}%
\setlength{\textfloatsep}{10pt}%
\setlength{\intextsep}{10pt}%
The \nextstate API is called by a search core, which defines the search strategy. For example, in \cref{lst:dmc:core} we define a search core that simply pops a \code{StateID} from a queue \code{Q} and requests the next states of the model \code{M} until the queue is empty. An example of such a model \code{M} is shown in \cref{lst:dmc:countermodel}, modeling four counters, going from 0 to 9, looping back to 0.
With each invocation of \code{nextStates()}, four next states are generated. For each \code{Slot} 0--3 in the state (line 5), the current value is obtained (line 6), it is incremented modulo 10 (line 7) and this delta is communicated to the model checker (line 8) using the \code{delta()} interface.
Thus, the initial state is \vectorbox{white}{0000} and for example
\vectorbox{white}{0974} generates the next states \vectorbox{white}{1974}, \vectorbox{white}{0074}, \vectorbox{white}{0984} and \vectorbox{white}{0975}.
The search core wraps the \dtree interface to add new states to the queue \code{Q}.
An example of this is shown in \cref{lst:dmc:deltaimpl}. Line 2 inserts the potentially new state in \dtree and is returned a \code{StateID} and an indicator if the state was new. If the state \code{s} is new and a root state (we will see a use case for non-root states in \cref{sec:dmc:example}), it is added to the queue \code{Q} in line 3. The \code{insert()} interface is implemented similarly, updating \code{Q} if needed. The \code{get()} interfaces are simply passed on.

\setlength{\floatsep}{1pt}%
\setlength{\textfloatsep}{5pt}%
\setlength{\intextsep}{1pt}%





\subsection{An example using a tree of states}
\label{sec:dmc:example}

\begin{wrapfigure}[9]{r}{.4\textwidth}
\centering%
\def\svgwidth{.7\columnwidth}
\scalebox{.54}{\ttfamily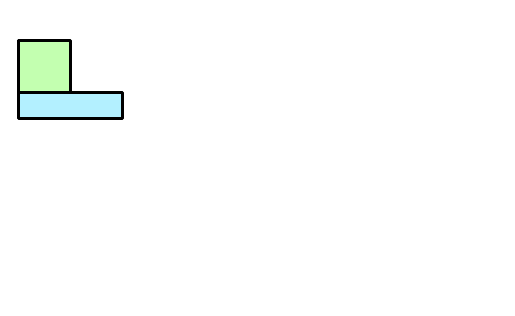}
\caption{%
A tree of states.
}%
\label{fig:dtree:processes}%
\end{wrapfigure}

\begin{mylisting}[!b]
\caption{Four processes counting from 0 to 9 and wrapping.}%
\label{lst:dtree:processes}%
\centering%
{\lstset{language=pseudo,style=pseudospecific}
\begin{lstlisting}
struct Process
IL^int pc;      # program counter
IL>int i;       # some variable
struct SV
IL^int n;       # number of processes
IL>StateID p[];
void nextStates(StateID rootID)
IL^SV sv = MC.get(rootID, true)                           # root state
IL|for(p in [0..sv.n])
IL|IL^int pi = MC.get(sv.p[p], &Process::i, 1, false)
IL|IL|pi = (pi + 1) % 10;
IL|IL|int newP = MC.delta(sv.p[p], &Process::i, {pi}, false);
IL|IL|int offset = &SV::p[p]                 # offset to p[p] within SV
IL>IL>MC.delta(rootID, offset, {newP}, true);              # root state
void initialState(StateID rootID)
IL^Process p = {1,0}                                 # pc is 1, i is 0
IL|StateID initP = MC.insert(p, false)
IL|SV rootState = {4, initP, initP, initP, initP}
IL>MC.insert(rootState, true)                             # root state
\end{lstlisting}}
\end{mylisting}

We can use the basic building blocks of uploading and downloading states for a more interesting concept: \emph{tree-structured states}.
The basic idea is that a state can contain a \code{StateID} that is associated with a non-root state, a sub-state. These are called sub-states because they could well be states of processes of which the combined state is the root state.

To illustrate this, we constructed a small model using processes in \cref{lst:dtree:processes}, which conceptually does the same as the model in \cref{lst:dmc:countermodel}. 
\Cref{fig:dtree:processes} illustrates what a complete state looks like. The important lines are 10--12, where we obtain the value of \code{i} of the current process (10), change it (11) and create a new sub-state with the changed value (12). Note that the \code{get()} in line 8 has \code{root=true}, indicating that the specified \code{StateID} is a root state, while the \code{get()} in line 10 does not. This matches with the corresponding \code{insert()}s and \code{delta()}s.

This is much like the \collapse method of SPIN~\cite{Holzmann97statecompression}
as briefly touched on in \cref{sec:intro:relatedwork}. SPIN supports a varying number of processes by having exactly this kind of structure: a root state with indices to indicate the state of individual processes. Just as \collapse, this allows to leverage the fact that duplicate processes share states. Since these states are inserted separately instead of a single, large vector, their data aligns, which helps in identifying common sub-vectors. However, \dtree supports arbitrary deep hierarchies of nested states.

\setlength{\floatsep}{1pt}%
\setlength{\textfloatsep}{10pt}%
\setlength{\intextsep}{10pt}%
\section{Experiments}
\label{sec:experiments}

To be able to compare to \treedbs, we have two options: 1)~initialise \treedbs with the length of the largest state and pad others with 0's; 2)~initialise \treedbs for the most common length and use a different storage component for other lengths. We implemented both options we call \treedbspad and \treedbsx{$S$}, where $S$ is the other storage component. Without these measures, \treedbs would not support variable-length states and could not run the variable-length experiments.

While we compared to \code{std::unordered\_map} of the STL library surrounded by an \code{std::mutex}, and a concurrent chaining hash map (cchm)~\cite{c71e99bcd01f48b99fdff1c495588533}, we omit the results for them as in all cases they perform roughly an order of magnitude worse. This is not unexpected, since they store an order of magnitude more memory because they do not compress the inserted states. The hash map \code{std::unordered\_map} specifically has a global mutex around it, causing high contention.
We could not make \compacttree into a DMC storage component because it needs a 64-bit index and the DMC API reserves only 40 bits for that purpose.

In addition to comparing these storages to each other, we compare DMC to the multicore implementation of LTSmin 3.0.2 with both \treedbs and \compacttree, both parallel DFS and BFS, using 64 threads. We also compare to SPIN 6.5.1 using \collapse and parallel BFS with 63 threads, as that is SPIN's maximum.
This gives us a baseline of where the performance of our model checker DMC as a whole is.

We ran DMC at 64 threads for a fair comparison, but it can use any number of threads. We tried both the DFS and BFS strategy for LTSmin, but these showed only minor difference. We will show the results for BFS, as we run SPIN using BFS as well and DMC currently only has a BFS search core.

The hardware that we ran our experiments on is ``caserta'', a Dell R930 with 2TiB of RAM and four E7-8890-v4 CPUs.
Each CPU has 24 cores, 60MiB of L3 cache and 512GiB of RAM, offering 96 physical cores in total. We ran our experiments on Linux 4.4.0 and all tools were compiled using GCC 9.3.0.

\subsection{Variable-length state models}

The design of \dtree is meant for software that uses dynamic memory. To test this we implemented three models: 1) a model that implements a concurrent hash map with a number of inserts~\cite{c71e99bcd01f48b99fdff1c495588533}; 2) a model of the concurrent Michael Scott queue~\cite{DBLP:conf/podc/MichaelS96} with various enqueue (E) and dequeue (D) operations; 3) a model of a sorted linked list (SSL).
All these data structures are modeled by creating a model that inserts a number of elements into them. The initial state is constructed by creating a number of processes, like in the example in \cref{sec:dmc:example}. These processes then run, with every interleaving explored, akin to how every interleaving of incrementing one of the four counters in the example of \cref{sec:dmc:example} is explored. For the Michael-Scott queue, we made a model that runs 3 enqueue operations and 3 dequeue operations in parallel. The Sorted Linked List model models a number of processes (6) that each insert a single element of 48 bytes (12 slots), dynamically allocated. The hash map model similarly inserts 9 pointers to elements of 16 slots in parallel. Dynamic memory is modeled using a \code{StateID} at the beginning of the root state to a memory slab sub-state that can expand.

\begin{figure*}[b!]
\centering
\hfill
\begin{subfigure}{.3\textwidth}
\centering%
\includegraphics[width=\textwidth]{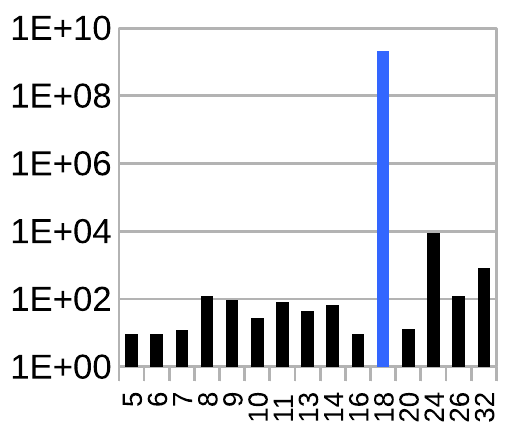}%
\vspace{-3pt}%
\caption{MSQ 3E+3E}%
\end{subfigure}\hfill
\begin{subfigure}{.3\textwidth}
\centering%
\includegraphics[width=\textwidth]{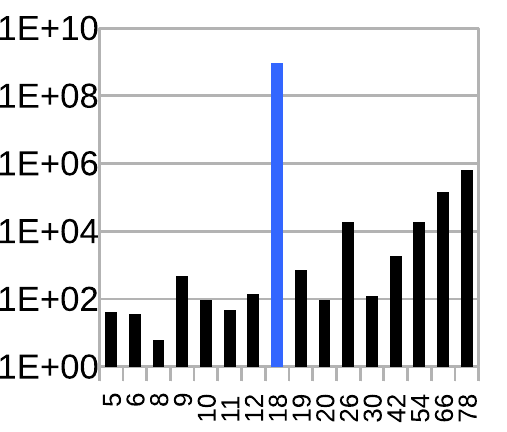}%
\vspace{-3pt}%
\caption{SSL (6)}%
\end{subfigure}\hfill
\begin{subfigure}{.3\textwidth}
\centering%
\includegraphics[width=\textwidth]{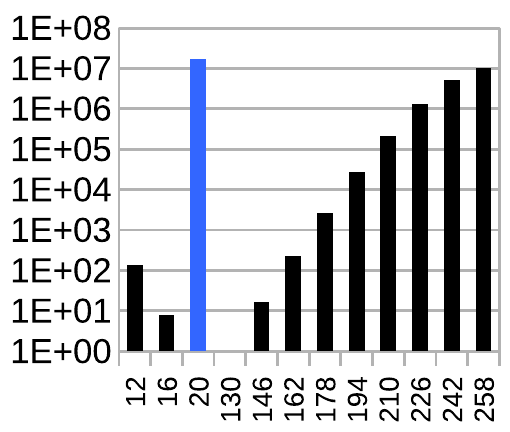}%
\vspace{-3pt}%
\caption{Hashmap 8/64}%
\end{subfigure}
\hfill${}$
\caption{The number of states (y-axis) inserted of a certain length in \code{\bfseries{}int}s (x-axis). The length of the root state is highlighted.}%
\label{fig:experiments:histograms}%
\end{figure*}

The effect these dynamic allocations and parallel insertions have on the distribution of the length of states is shown in \cref{fig:experiments:histograms}. For example, the hash map model is most pronounced: starting with an initial global memory of 130 slots (some string data), the first process dynamically allocates 16 slots and inserts the element, yielding a number of states of length 146. If we add a second process and all the interleavings, we get many more states of length 162, etc.

{\rowcolors{2}{black!5!white}{black!10!white}%
\setlength{\extrarowheight}{2pt}%
\renewcommand{\arraystretch}{1.0}%
\begin{table}[!b]%
\caption{Experiments that use the \dmcapi. Set scale X-Y(-Z) means $2^X$ root set nodes, $2^Y$ data set nodes, $2^Z$ entries for the sub-storage. \treedbsx{cchm}($L$) and \treedbspad($L$) mean they are initialized for states of length $L$ (in \code{\bfseries{}int}s).%
}%
\label{tab:experiments:results}%
\centering%
\scalebox{.95}{\begin{tabular}{cllrr}%

\rowcolor{darkgray}
\headcell{Model}
& \headcell{Storage}
& \headcell{Set Scale}
& \headcell{Time}
& \headcell{B/state}
\\

\cellcolor{black!15!white}
& \treedbsx{cchm}(18)
& 32-30-24
& 4917.33s
& 8.03
\\

\cellcolor{black!15!white}
& \treedbspad(32)
& 32-30
& 469.86s
& 9.16
\\

\cellcolor{black!15!white}\multirow{-3}{.6cm}{\begin{sideways}\vphantom{q}MSQ\end{sideways}\begin{sideways}\small{3E+3D}\end{sideways}}
& \dtree
& 32-30
& 392.45s
& 9.16
\\
\hline

\cellcolor{black!15!white}
& \treedbsx{cchm}(20)
& 34-32-24
& 2497.33s
& 8.97
\\

\cellcolor{black!15!white}
& \treedbspad(78)
& 34-32
& 777.18s
& 24.86
\\

\cellcolor{black!15!white}\multirow{-3}{.6cm}{\begin{sideways}\scriptsize{}\vphantom{q}\hphantom{HH}SLL\end{sideways}\begin{sideways}\scriptsize{}{(6 inserts)}\end{sideways}}
& \dtree
& 34-32
& 416.22s
& 12.89
\\
\hline

\cellcolor{black!15!white}
& \treedbsx{cchm}(18)
& 28-28-28
& 187.01s
& 1072.95
\\

\cellcolor{black!15!white}
& \treedbspad(258)
& 28-28
& 65.18s
& 75.62
\\

\cellcolor{black!15!white}\multirow{-3}{.6cm}{\begin{sideways}\scriptsize{}Hashmap\end{sideways}\begin{sideways}\scriptsize{}{(8 inserts)}\end{sideways}}
& \dtree
& 28-28
& 24.01s
& 55.00
\\
\hline

\cellcolor{black!15!white}
& \treedbsx{cchm}(18)
& 34-32-30
& 21171.64s
& 1137.57
\\

\cellcolor{black!15!white}
& \treedbspad(274)
& 34-32
& 1391.79s
& 75.62
\\

\cellcolor{black!15!white}\multirow{-3}{.6cm}{\begin{sideways}\scriptsize{}Hashmap\end{sideways}\begin{sideways}\scriptsize{(9 inserts)}\end{sideways}}
& \dtree
& 34-32
& 477.25s
& 54.30
\\

\end{tabular}}%
\end{table}}

The results for variable-length state models are shown in \cref{tab:experiments:results}.
Overall, \dtree is clearly the faster of the three storages. It is up to 44 times faster than \treedbsx{cchm} and 1.2--2.9 times faster than \treedbspad. The slow times of \treedbsx{cchm} are largely caused by the slower, uncompressed cchm that is used for states of lengths other than the root state so the results have to be interpreted as such. For \treedbspad, we see the downside of padding with zeroes: with increased state-length variance, the overhead increases and performance drops.

Regarding compression, we notice \treedbsx{cchm} is actually better for the MSQ and Sorted Linked List models. This can be explained by looking at the variance of the distribution of state lengths: the root state length (18) dominates all other lengths. The shape of the tree of nodes of \treedbs has 2 nodes on the lowest level and 7 on the second-to-lowest. \Dtree on the other hand, has 8 nodes on the deepest level and 1 on the second-to-highest, because it uses a chain of balanced trees, in this case leading to 16 state slots (left) and 2 (right).
As theorised in \cref{sec:dtree:chain}, this difference causes that \dtree often needs \emph{one more node} for even a small change. Since often the memory is changed, which is modeled by a \code{StateID} at the beginning of the root state, this is precisely what happens.
Thus, the root state length is so dominant that the overhead of uncompressed cchm entries is less than the overhead of the less balanced tree that \dtree uses.

When we look at the Hashmap model, which has a significantly higher variance, we see that \treedbsx{cchm} requires an order of magnitude more space. Here, \treedbs stores only states of length 20 and all others are stored in the sub-storage cchm, which does not compress states.

In general these results show that dedicated support for variable-length outperforms padding zeroes (\treedbspad) and offloading other-sized vectors (\treedbsx{cchm}). The results also show that an increase of variance in state-length increases the advantage of \dtree.

\subsection{Fixed-length state models}
\label{sec:experiments:fixed}
\begin{figure*}[!b]
\begin{subfigure}{.32\textwidth}
\centering%
\begin{overpic}[width=\textwidth]{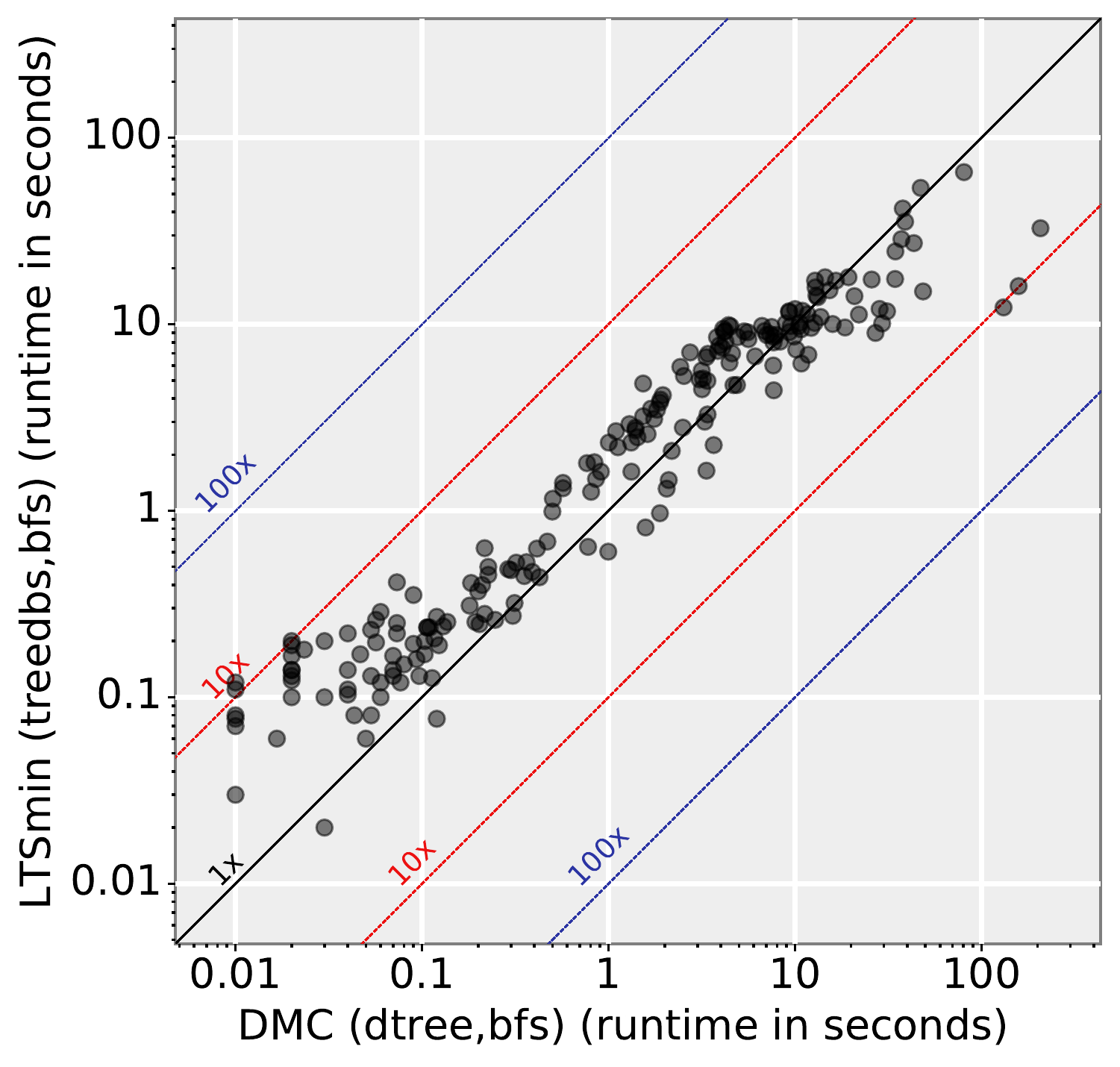}%
\put(17,83){\phantomsubcaption\label{fig:experiments:beemresults:vstreetime}\large{(\subref*{fig:experiments:beemresults:vstreetime})}}%
\end{overpic}%
\vspace{-3pt}%
\end{subfigure}
\begin{subfigure}{.32\textwidth}
\centering%
\hspace{-7pt}\begin{overpic}[width=\textwidth]{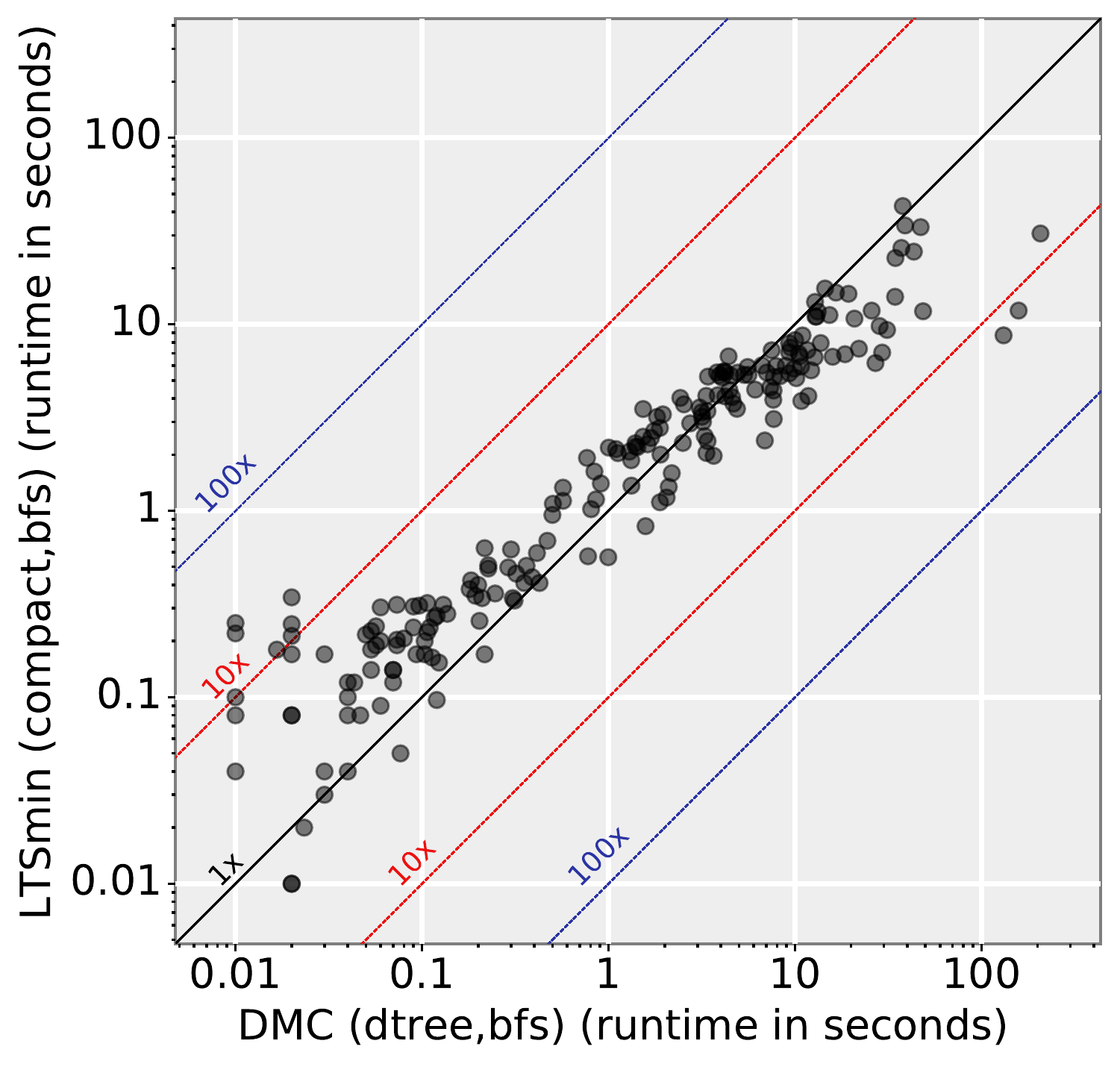}%
\put(17,83){\phantomsubcaption\label{fig:experiments:beemresults:vscompacttime}\large{(\subref*{fig:experiments:beemresults:vscompacttime})}}%
\end{overpic}%
\vspace{-3pt}%
\end{subfigure}
\begin{subfigure}{.32\textwidth}
\centering%
\hspace{-17pt}\begin{overpic}[width=\textwidth]{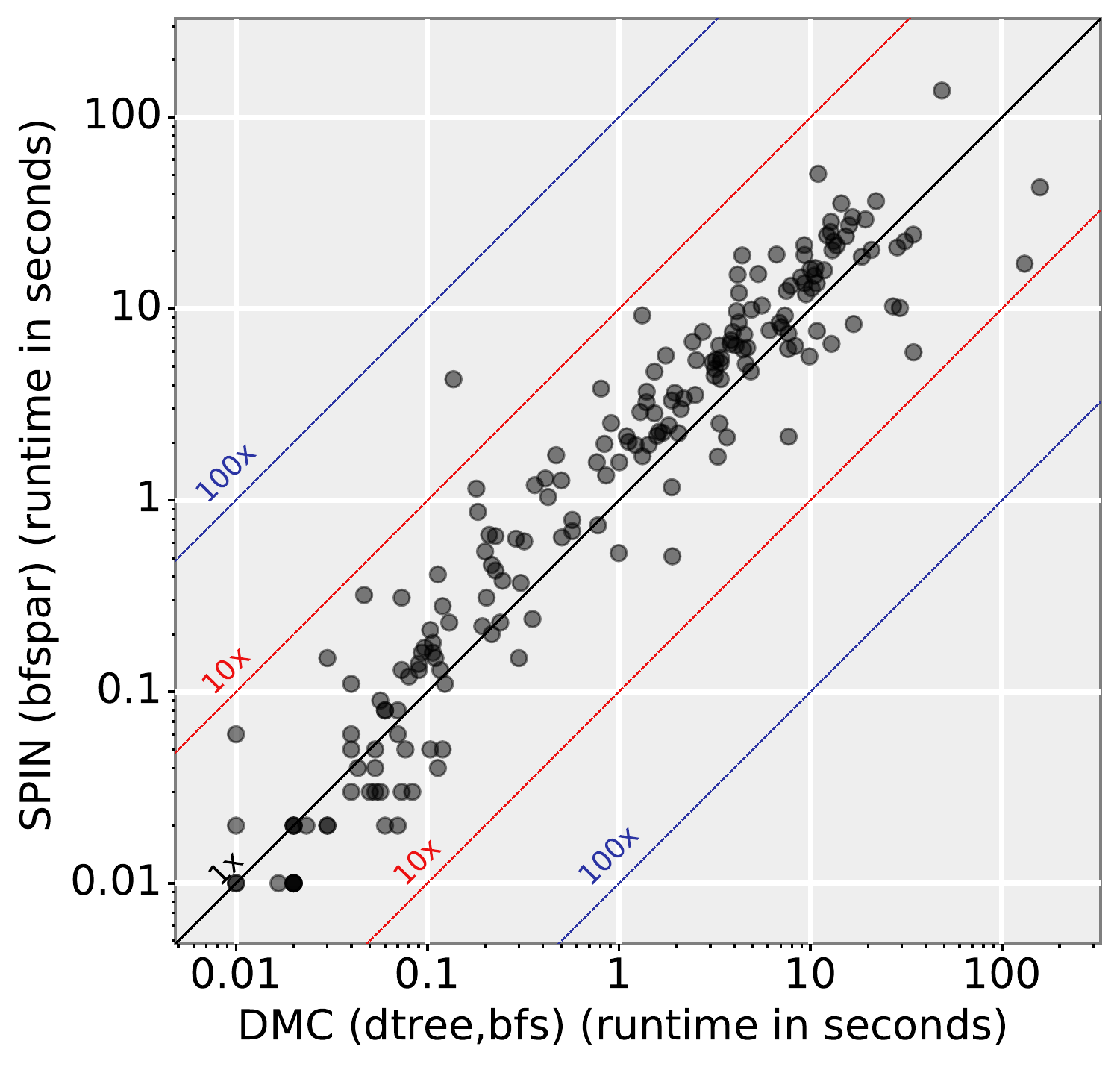}%
\put(17,83){\phantomsubcaption\label{fig:experiments:beemresults:vsspintime}\large{(\subref*{fig:experiments:beemresults:vsspintime})}}%
\end{overpic}%
\vspace{-3pt}%
\end{subfigure}
\begin{subfigure}{.32\textwidth}
\centering%
\begin{overpic}[width=\textwidth]{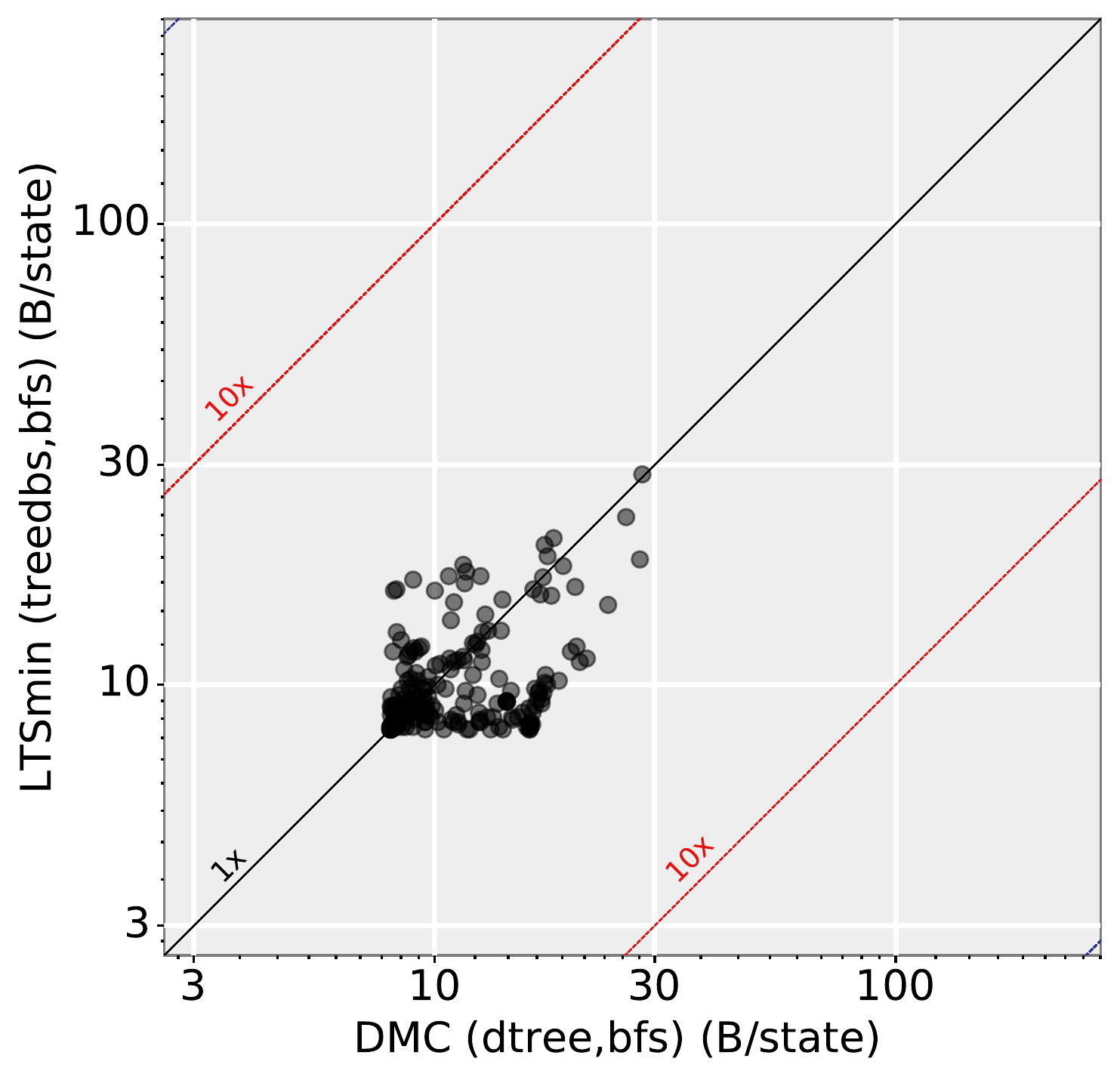}%
\put(15,85){\phantomsubcaption\label{fig:experiments:beemresults:vstreebytesperstate}\large{(\subref*{fig:experiments:beemresults:vstreebytesperstate})}}%
\end{overpic}%
\vspace{-3pt}%
\end{subfigure}
\begin{subfigure}{.32\textwidth}
\centering%
\begin{overpic}[width=\textwidth]{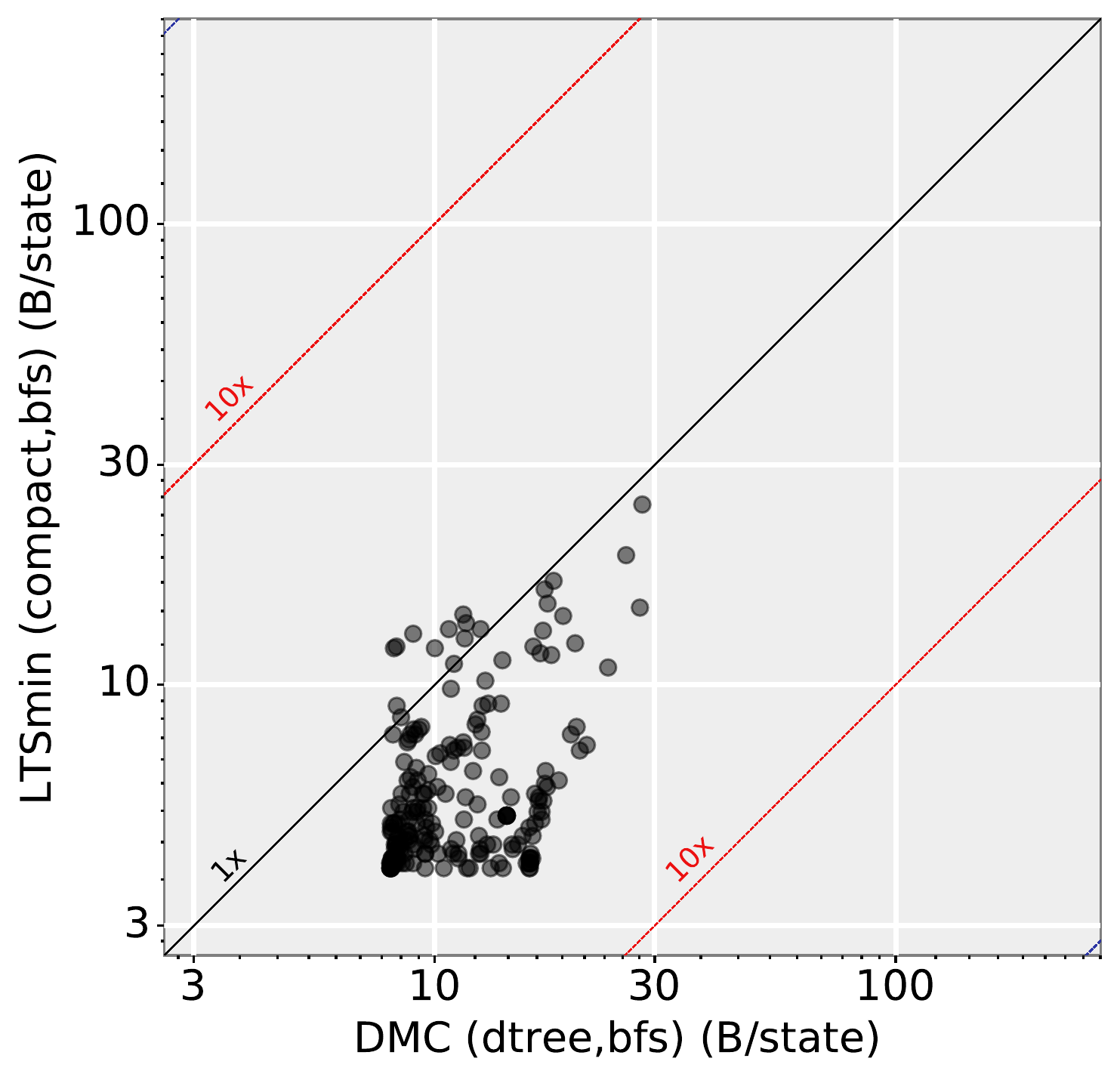}%
\put(15,85){\phantomsubcaption\label{fig:experiments:beemresults:vscompactbytesperstate}\large{(\subref*{fig:experiments:beemresults:vscompactbytesperstate})}}%
\end{overpic}%
\vspace{-3pt}%
\end{subfigure}
\begin{subfigure}{.32\textwidth}
\centering%
\begin{overpic}[width=\textwidth]{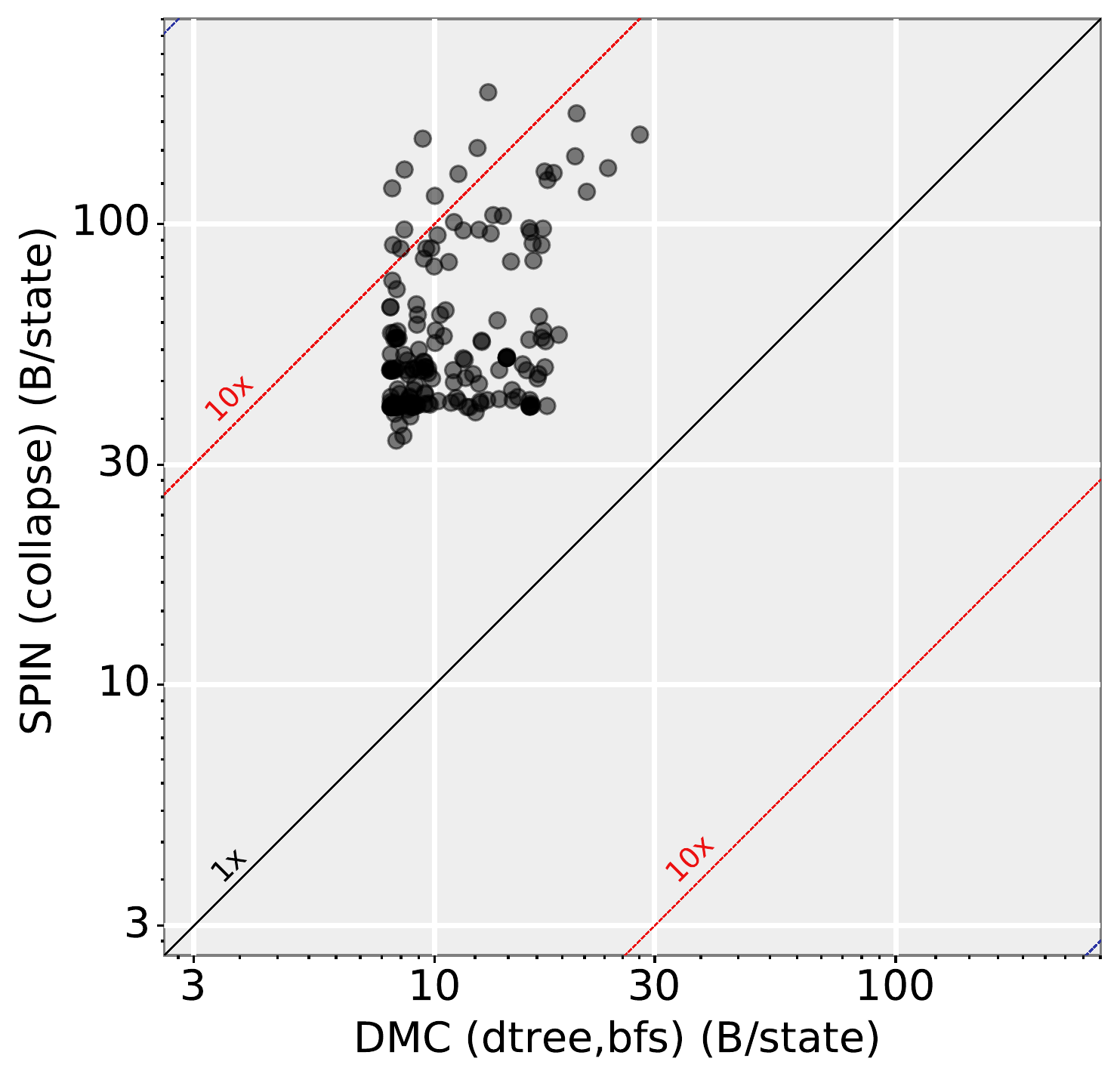}%
\put(15,85){\phantomsubcaption\label{fig:experiments:beemresults:vsspinbytesperstate}\large{(\subref*{fig:experiments:beemresults:vsspinbytesperstate})}}%
\end{overpic}%
\vspace{-3pt}%
\end{subfigure}
\caption{\small{DMC/dtree compared to
LTSmin/\treedbs (\subref*{fig:experiments:beemresults:vstreetime},%
\subref*{fig:experiments:beemresults:vstreebytesperstate}),
LTSmin/compact (\subref*{fig:experiments:beemresults:vscompacttime},%
\subref*{fig:experiments:beemresults:vscompactbytesperstate})
and SPIN (\subref*{fig:experiments:beemresults:vsspintime},%
\subref*{fig:experiments:beemresults:vsspinbytesperstate}) on runtime
(\subref*{fig:experiments:beemresults:vstreetime},%
\subref*{fig:experiments:beemresults:vscompacttime},%
\subref*{fig:experiments:beemresults:vsspintime}) and bytes per state
(\subref*{fig:experiments:beemresults:vstreebytesperstate},%
\subref*{fig:experiments:beemresults:vscompactbytesperstate},%
\subref*{fig:experiments:beemresults:vsspinbytesperstate})}.%
}
\label{fig:experiments:beemresults}%
\end{figure*}

To evaluate the cost of adding variable-length support, we also run experiments using models with fixed-length states.
We modified SpinS~\cite{eemcs22042} to emit models that implement the API of DMC for models from the BEEM database~\cite{10.1007/978-3-540-73370-6_17}. For SPIN, these states could be of varying length, but SpinS emits fixed-length states models since its primary target is PINS for LTSmin. Thus, these tests do not use the \code{delta()} interface, but only insert and get complete states. Of these models, we include the results for 118 models. Other models could not be compared because either they have two adjacent state slots with -1 values (\treedbs and \compacttree do not support that, e.g.\ \textsc{gear.1}, in which case LTSmin aborts), the \promela could not be translated (e.g. \textsc{train-gate.1}) or they are too large for all tools to finish (e.g. \textsc{public\_subscribe.5}).

{\rowcolors{2}{black!5!white}{black!10!white}%
\setlength{\extrarowheight}{1pt}%
\renewcommand{\arraystretch}{0.95}%
\begin{wraptable}[11]{r}{.44\textwidth}
\centering%
\scalebox{.8}{\begin{tabular}{lR{1.5cm}R{1.5cm}}%
\rowcolor{darkgray}
\headcell{Tool}
& \headcell{Total Time (s)}
& \headcell{Average B/state}
\\

SPIN (collapse)
& 6081
& 71.8
\\

SPIN (bfspar)
& 1050
& 146
\\

DMC/\treedbsx{cchm}
& 780
& 8.98
\\

DMC/\treedbspad
& 773
& 8.98
\\

DMC/\treedbsx{stdmap}
& 749
& 8.98
\\

DMC/dtree
& 748
& 11.7
\\

LTSmin/treedbs
& 598
& 8.98
\\

LTSmin/compact
& 428
& 4.98
\\

\end{tabular}}%
\caption{Totals for 118 models from the BEEM database.}%
\label{tab:experiments:summaries}%
\end{wraptable}%

\setlength{\floatsep}{0pt}%
\setlength{\textfloatsep}{0pt}%
\setlength{\intextsep}{0pt}%

The results for the comparison to LTSmin and SPIN using models from the BEEM database are shown in \cref{fig:experiments:beemresults,tab:experiments:summaries}. When looking at \cref{fig:experiments:beemresults:vstreetime}, we can see that the combination of \dmc with \dtree is nearing the performance of LTSmin with \treedbs time-wise. There are a number of outliers in both directions, but in the more time-consuming models one can see that LTSmin with \treedbs has the edge over DMC, still. In terms of bytes per state (\cref{fig:experiments:beemresults:vstreebytesperstate}) we see a similar result. This shows in the total runtime and B/states as well: LTSmin/\treedbs takes 20\% less time and 23\% less memory on average.

\setlength{\floatsep}{8pt}%
\setlength{\textfloatsep}{10pt}%
\setlength{\intextsep}{10pt}%

The comparison to \compacttree follows the same trend time-wise (\cref{fig:experiments:beemresults:vscompacttime}). \Compacttree clearly outperforms \dtree in terms of bytes per state (\cref{fig:experiments:beemresults:vscompactbytesperstate}). The use of a Cleary table for the root set is a clear winner.

\Cref{fig:experiments:beemresults:vsspintime} compares DMC/dtree to SPIN with parallel BFS. Here, we see that DMC/dtree has the edge over SPIN, time-wise. \Cref{fig:experiments:beemresults:vsspinbytesperstate} compares DMC/dtree to SPIN with \collapse, but without parallel BFS (these cannot be used simultaneously). Even then, SPIN with \collapse is outperformed by DMC/dtree in all benchmarks.

Comparing the storage components of DMC between themselves (\cref{tab:experiments:summaries}), we notice all \treedbs variants compress equally and also equal to LTSmin/\treedbs. This is expected, as for fixed-length states, the implementations should be equivalent.
Furthermore, we see that \dtree loses in terms of bytes per state. The different structure of compression tree is again a likely cause: using the chain of perfectly balanced trees on average results in a less balanced tree than \treedbs and thus on average a change (via \code{insert()} or \code{delta()}) requires more new nodes.
This is the price to pay to support variable-length states.

\section{Discussion and conclusion}

We presented \dtree, a concurrent variable-length state storage component that stores states in a compression tree. It allows partial reconstruction of states and incremental updates to parts of states \emph{without reconstructing the entire states}. To showcase \dtree, we implemented the DMC Model Checker
such that we can expose the functionality of \dtree through the \dmcapi.
We implemented three variable-length state vector models that make full use of the \dmcapi.

We compare \dtree to other state storage components, such as \treedbs. In the case of \treedbs, we \emph{pad} all vectors to the length of the largest vector (manually determined). Here, \dtree is able to perform up to 2.9 times faster than \treedbs using 29\% less memory on average, without the a priori need to manually determine the largest vector. The advantage of \dtree increases as the difference in length between states increases.

To evaluate how much efficiency is lost to gain variable-length support, we compare \dmc using \dtree to LTSmin and SPIN using models from the BEEM database with fixed-length states. In this setting, \dtree is actually marginally faster than \treedbs, but uses 30\% more memory.
\Dmc as a whole can only approach the performance of fixed-length state model checker LTSmin, sacrificing 20--23\%. It is still outclassed by \compacttree in terms of bytes per state, which compressed up to 2.3x more than \dtree. In the same tests, \dmc with \dtree is faster and provides a higher compression than SPIN with \collapse.
Compared to SPIN using \collapse, \dtree is 8.1x faster and uses 6.1x less memory.
Even though \dmc nor \dtree are particularly optimized for fixed-length complete-state changes, the performance overall is reasonable.

This research is part of the ongoing research towards creating a software model checker for use in a continuous integration pipeline. With the advent of \dtree\footnote{The source code of \dtree can be found at \url{https://github.com/bergfi/dtree}.}, we are now one step closer.

\subsection{Future work}

We have seen that \compacttree outperforms \dtree for fixed-length states. \Compacttree uses a Cleary Table for its root set. To improve \dtree, we can investigate if we can leverage such a table, since \compacttree as is uses a 64-bit index and the DMC API currently has only 40 bits available for the state index.

DMC still needs to improve as well. For example, the search core of DMC is a simple parallel BFS, lacking a more sophisticated work-stealing algorithm. Implementing such a feature would improve the performance of DMC as a whole.

We aim to use DMC as the core for our upcoming multi-core software model checker LLMC. The purpose of LLMC is to model check LLVM IR assembly code. To model the stack and heap, \dtree lends itself perfectly. We can then compare the resulting implementation with \divine.

\subsection{Acknowledgements}
The author would like to thank Arnd Hartmanns and Jaco van de Pol for their invaluable contributions and Alfons Laarman for discussions on \treedbs and \compacttree.
This research is sponsored by 3TU Big Software on the Run project (\url{http://www.3tu-bsr.nl/}).

\clearpage
\bibliographystyle{IEEEtran}
\bibliography{common/bib}

\end{document}